\documentstyle[12pt]{article}

\def\tree{{\rm tree}}
\def\oneloop{{1 \mbox{-} \rm loop}}

\def\pol{\varepsilon}
\def\qb{{\bar q}}
\def\Qb{{\overline Q}}

\def\ib{{\bar\imath}}
\def\jb{{\bar\jmath}}

\def\Tr{\mathop{\rm Tr}\nolimits}

\def\spa#1.#2{\left\langle#1\,#2\right\rangle}
\def\spb#1.#2{\left[#1\,#2\right]}
\def\L{\left(}

\def\L#1#2{\left\{#2\right\}_{#1}}

\def\eqn#1{eq.~(\ref{#1})}

\def\Eqns#1#2{eqs.~(\ref{#1}) - (\ref{#2})}

\def\sec#1{section~{\ref{#1}}}

\def\tab#1{table~\ref{#1}}
\def\Tab#1{Table~\ref{#1}}

\def\feynsl#1{
  \setbox0=\hbox{/} \setbox1=\hbox{$#1$}
  \dimen0=\wd0 \advance\dimen0 by -\wd1 \divide\dimen0 by 2
  \ifdim\wd0>\wd1 \raise.15ex\copy0\kern-\wd0\kern\dimen0\copy1\kern\dimen0
  \else \kern-\dimen0\raise.15ex\copy0\kern-\dimen0\kern-\wd1\copy1\fi}

\newskip\humongous \humongous=0pt plus 1000pt minus 100pt

\newif\ifdtup

\makeatletter
\def\@eqnnum{\hbox{\reset@font\rm(\theequation)}}
\let\make@eqnnum=\@eqnnum %
\def\eqnum#1{\dec@eqnnum \global\def\make@eqnnum{\reset@font\rm(#1)}%
\def\@currentlabel{#1}%
}
\def\inc@eqnnum{\addtocounter{equation}{1}}
\def\dec@eqnnum{\addtocounter{equation}{-1}}
\@definecounter{equation}%
\@addtoreset{equation}{section} %
\def\theequation@prefix{{\thesection}.} %
\def\theequation{\theequation@prefix\arabic{equation}}%
\makeatother

\def\spa#1.#2{\left\langle#1\,#2\right\rangle}
\def\spb#1.#2{\left[#1\,#2\right]}
\def    \br#1#2{\langle#1\,#2\rangle}
\def    \sq#1#2{[  #1 \, #2 ]}
\def    \sap#1#2#3{{\langle #1 | #2 |#3  \rangle} }
\def    \t#1#2#3{{t_{#1 #2 #3}} }
\def    \s#1#2{{s_{#1 #2}} }
\def    \sapp#1#2#3#4{{\langle #1 | (#2+#3) |#4  \rangle} }

\begin{document}
\begin{titlepage}

\hspace*{\fill}\parbox[t]{4cm}{
BNL-HET-99/27\\
DFTT 50/99\\
\today}

\begin{center}
{\Large\bf Multi-Photon Amplitudes\\ for Next-to-Leading Order QCD} \\
\vspace{1.cm}

{Vittorio Del Duca}\\
\vspace{.2cm}
{\sl Istituto Nazionale di Fisica Nucleare\\ Sezione di Torino\\
via P. Giuria, 1\\ 10125 - Torino, Italy}\\

\vspace{.8cm}
{William B. Kilgore}\\
\vspace{.2cm}
{\sl Department of Physics\\
Brookhaven National Laboratory\\
Upton, New York 11973-5000, USA}\\

\vspace{.5cm}
and \\
\vspace{.5cm}

{Fabio Maltoni}\\
\vspace{.2cm}
{\sl Dipartimento di Fisica Teorica\\
Universit\`a di Torino\\
via P. Giuria, 1\\ 10125 - Torino, Italy}\\

\vspace{.5cm}

\begin{abstract}
We present the tree-level amplitudes involving one, two and three
photons that are required for next-to-leading order QCD calculations
of production rates of three final-state particles. We also present
the required one-loop amplitudes in terms of previously published
results.
\end{abstract}

\end{center}
 \vfil

\end{titlepage}

\section{Introduction}

The primary goal of the LHC physics program is the investigation of
the mechanism of electroweak symmetry breaking; to observe the Higgs
boson, if it exists, and to study its properties.  If the mass of the
Higgs is light (100 GeV $\le m_H \le$ 140 GeV), studies indicate that
the rare decay into two photons, $H\to\gamma\gamma$ provides the best
signature \cite{LHCdocs}.  Because the signal-to-background ratio will
be quite low ($\sim 7\%$)~\cite{LHCdocs}, this promises to be a
demanding analysis.  Unfortunately, our theoretical
understanding of both the signal and the background is
less than satisfactory.  The next-to-leading order (NLO) QCD
corrections to Higgs production are known to be very large
(${\cal O}(100\%)$)\ \cite{zerwas}.  The NLO corrections to the QCD
background~\cite{owens}, $pp\to\gamma\gamma$, (for which the full NLO
fragmentation contributions are just now being evaluated~\cite{pilon})
are known to also be very large, and there are strong indications that
the next-to-next-to-leading order (NNLO) corrections will also be
sizable because of the large gluon luminosity at LHC which will
enhance the $gg\to\gamma\gamma$ subprocess~\cite{kunszt,yuan} that
first appears at NNLO.  Thus it seems that we will need (at least)
full NNLO calculations of both the signal and the background in order
to obtain a reasonable theoretical understanding of this process.

In order to improve the signal-to-background ratio, Higgs production
in association with a high transverse energy ($E_T$) jet, $pp\to H\
jet \to \gamma\gamma\, jet$, has been considered~\cite{abdullin}.
This process offers the advantage of being more flexible with respect
to choosing suitable acceptance cuts to curb the background.  The
$pp\to H\ jet$ process is known to leading order (LO)
exactly~\cite{ellis}, while the NLO calculation~\cite{grazz} has been
performed in the infinite top mass limit.  The NLO corrections to the 
signal are also large, but it is believed that the background,
$pp\to\gamma\gamma\, jet$, can be more reliably calculated 
because LO production is dominated by the $qg\to q\gamma\gamma$
sub-process, which benefits from the large gluon luminosity, while
the $gg\to g \gamma\gamma$ sub-process, which is believed to dominate
the NNLO contribution, yields a comparatively small
contribution~\cite{kunszt}.  Thus an evaluation of the full NLO
corrections to the background should provide a reliable quantitative
estimate.

In order to evaluate the NLO corrections to $pp\to\gamma\gamma\, jet$,
we need tree-level QCD amplitudes with six external legs ($\qb
qgg\gamma\gamma$, $\qb q\Qb Q\gamma\gamma$) and one-loop amplitudes
with five external legs ($\qb qg\gamma\gamma$). 
The same amplitudes also contribute to the evaluation of the NNLO
corrections to $pp\to\gamma\gamma$ scattering.
Of course, there are other processes involving one or more photons that can be
computed at NLO, including the
recently observed direct photo-production of three jets at
HERA~\cite{zeus}.  

In principle, the amplitudes required for the calculations
mentioned above can be
obtained from pure QCD amplitudes by turning external gluons into
photons.  This is done by summing over permutations of the
color-ordered amplitudes~\cite{Color} to remove the non-Abelian
character of the gluon coupling (see
secs.~\ref{sec:twotwo},\ref{sec:qqgggamma}).  This is the procedure we
use to obtain the one-loop amplitudes.  Of course, this procedure
naturally results in more complicated expressions than those for the
pure QCD amplitudes with which one starts.  One generally obtains far
more compact expressions by computing the photon amplitudes directly,
and this is what we do for the tree-level amplitudes.

In this paper, we present all of the matrix
elements needed to compute at NLO two-parton to three-parton
scattering where one, two or three of the partons are photons. 
These amplitudes, with the one-loop amplitudes supplemented by
higher-order terms in the dimensional-regularization parameter
$\epsilon$~\cite{us}, also contribute to the computation at NNLO of
two-parton to two-parton scattering where one or two of the partons
are photons. Many
of these amplitudes have been discussed previously.  The five-parton
tree-level amplitudes involving photons have been known for some
time~\cite{wu,mpReview}.  The five-parton one-loop amplitudes
involving one photon have been discussed in detail~\cite{BDK,AS} and 
are reproduced here for completeness.  The new results presented
here are the compact expressions for the six-parton tree-level
amplitudes involving photons and the detailed expansions for the
five-parton one-loop multi-photon amplitudes.

The paper is organized as follows:
in section 2 we present the known results
for the amplitudes at tree level and explain the rules
which allow one to turn external gluons into photons. 
In sections 3, 4 and 5, we evaluate the six-parton tree-level amplitudes
with one, two and three photons, respectively.
In section 6, we present detailed expansions for the corresponding 
five-parton one-loop amplitudes in terms of known results.
In section 7 we draw our conclusions.

\section{Tree-level Amplitudes}
\label{sec:amps}

It is now common to express multi-parton amplitudes in QCD using a
color-ordered/helicity decomposition~\cite{Color}.  The usual
color-ordered decomposition follows from replacing the group structure
constants with commutators of fundamental representation matrices
\footnote{We normalize the fundamental representation matrices such
that $\Tr[T^aT^b] = \delta^{ab}$.}
\begin{equation}
f^{abc} = -{i \over \sqrt{2}} (\Tr[T^a T^b T^c] - \Tr[T^b T^a T^c])
\end{equation}
and then using $SU(N_c)$ Fierz identities
\begin{equation}
T^a_{i \jb} \, T^a_{m \bar n} =  \delta_{i \bar n} \,\delta_{m \jb}
  - {1\over N_c}\, \delta_{i \jb} \, \delta_{m \bar n}
\end{equation}
to combine traces.  Each distinct ordering of representation matrices
(Chan-Paton, or color factor) is associated with a gauge invariant
sub-amplitude depending only on the ordering of the parton labels.
The full amplitude is expressed as the sum over the independent color
factors multiplying their associated sub-amplitude.

The helicity decomposition is accomplished by specifying the helicity
of each external leg.  Typically, helicity amplitudes are written in
terms of spinor products $\spa{i}.j$, $\spb{i}.j$.  The spinor
products are antisymmetric, with norm $|\spa{i}.j| = |\spb{i}.j| =
\sqrt{|s_{ij}|}$.  We use the convention of Mangano and
Parke~\cite{mpReview}, defining 
$\spa{i}.j = \langle i^- | j^+\rangle$ and $\spb{i}.j = \langle i^+|
j^-\rangle$, where $|i^{\pm}\rangle$ are massless Weyl spinors of
momentum $k_i$, labeled with the sign of the helicity.  In this
convention, the spinor products obey the rule $\spa{i}.j\spb{j}.i =
s_{ij} = 2k_i\cdot k_j$.

\subsection{Two-quark amplitudes}
\label{sec:twoone}

Tree-level amplitudes involving two quarks and $(n-2)$ gluons are
written as~\cite{Color,mpReview},
\begin{equation}
 {\cal A}_n^\tree(\qb_1,q_2;g_3,\ldots,g_{n})
  = g^{n-2} \sum_{\sigma\in S_{n-2}}
   (T^{\sigma_3}\ldots T^{\sigma_n})_{i_2}^{~\ib_1}\
    A_n^\tree(\qb_1,q_2,g_{\sigma_3},\ldots,g_{\sigma_n})\, ,
\label{TwoQuarkGluonDecomp}
\end{equation}
where $S_{n-2}$ is the permutation group on the $n-2$ elements
$(3\dots n)$.  The dependence on parton helicities and momenta is
implicit.

For the {\sl maximally-helicity-violating} configurations, in which
two partons are of one helicity and all others are of the opposite
helicity (where helicity is defined as if all particles are outgoing),
say $(-,-,+,\dots,+)$, there is only one independent color/helicity
sub-amplitude
\begin{equation}
A_n^\tree(\qb_1^+, q_2^-, g_3,\dots, g_n) =
i{\spa1.i {\spa2.i}^3 \over \spa1.2\spa2.3 \cdots\spa{n}.1}
\label{dueb}
\end{equation}
%
where gluon $g_i$ has negative helicity.  (The quark and
anti-quark must have opposite helicity in order to conserve helicity
along the fermion line.)

All other color/helicity amplitudes can be obtained by relabeling and
by use of the discrete symmetries of parity inversion and charge
conjugation.  Parity inversion flips the helicities of all particles.
It is accomplished by the ``complex conjugation'' operation (indicated
with a superscript ${}^\dagger$) defined such that $\spa{i}.j
\leftrightarrow \spb{j}.i$, but explicit factors of $i$ are not
conjugated to $-i$. In addition, there is a factor of $-1$
for each pair of quarks participating in the amplitude.  Charge
conjugation changes quarks into anti-quarks without inverting
helicities.  In addition, there is a reflection symmetry in the color
ordering such that
\begin{equation}
A_n^\tree(\qb_1^+, q_2^-, g_3,\dots, g_n) = (-1)^n
A_n^\tree(q_2^-, \qb_1^+, g_n ,\dots, g_3)
\end{equation}

So, if gluon $i$ has negative helicity and all other gluons have
positive helicity, one finds the sub-amplitude $A_n^\tree(\qb_1^-,
q_2^+, g_3,\dots, g_n)$ to be
\begin{equation}
    A_n^\tree(\qb_1^-, q_2^+, g_3,\dots, g_n) = 
  (-1)^n A_n^\tree(\qb_2^+, q_1^-, g_n,\dots, g_3) =
 i{{\spa1.i}^3\spa2.i \over \spa1.2\spa2.3\dots\spa{n}.1}.
\end{equation}

\subsection{Turning gluons into photons}
\label{sec:twotwo}

The color decomposition of amplitudes involving two quarks, $m$ photons
and $r$ gluons, with $m+r=n-2$, is
\begin{eqnarray}
\lefteqn{ {\cal A}_n^\tree(\qb_1, q_2; g_3, \dots, g_{r+2};
\gamma_{r+3}, \dots, \gamma_{n}) = } \label{TwoQuarkGluonPhot}\\
 &&g^{r}\, (\sqrt{2}\, e Q_q)^m \,
\sum_{\sigma\in S_r}
   (T^{\sigma_3}\ldots T^{\sigma_{r+2}})_{i_2}^{\ib_1}\
A_n^\tree(\qb_1,q_2,g_{\sigma_3},\ldots,g_{\sigma_{r+2}},
\gamma_{r+3},\dots,\gamma_{n})\, .\nonumber
\end{eqnarray}
The color-ordered amplitudes can be computed directly or can be
obtained from two-quark $(n-2)$ gluon amplitudes by converting gluons
into photons.  In order to convert a gluon into a photon we replace
the quark-gluon vertex factor $gT^a$ with the quark-photon vertex
factor\footnote{The factor $\sqrt{2}$ is due to our choice of
normalization of the fundamental representation matrices.} $\sqrt{2}\,
e Q_qI$.  Since the identity matrix $I$ commutes with the $SU(N)$
matrices $T^a$, all possible attachments of the photon to the quark line
(leaving the gluon ordering fixed) contribute to the same color
structure.   That is, matrix elements involving photons are obtained
by summing over permutations of gluon matrix elements in which the
photon assumes all possible places in the ordering.  For example,
\begin{eqnarray}
&&A_{5}^\tree(\qb_1,q_2,g_3,g_4;\gamma_5) =\\
&&\phantom{A_5^\tree(\qb_1)}
  A_{5}^\tree(\qb_1,q_2,g_3,g_4,g_5) +
  A_{5}^\tree(\qb_1,q_2,g_3,g_5,g_4) + 
  A_{5}^\tree(\qb_1,q_2,g_5,g_3,g_4).\nonumber
\end{eqnarray}

For maximally-helicity-violating configurations, $(-,-,+,\dots,+)$,
the sub-amplitudes are found to be~\cite{Color,mpReview},
\begin {equation}
A_n^\tree(\qb_1^+, q_2^-, g_3,\dots, g_{r+2}, \gamma_{r+3}, \dots,
\gamma_{r+m+2}) = i{\spa1.i{\spa2.i}^3 \over \spa1.2\spa2.3 \cdots
\spa{(r+2)}.1}\, \prod_{j=r+3}^{r+m+2}\left[
{\spa2.1 \over \spa2.j\spa{j}.1}\right]
\label{mhvphotamp}
\end{equation}
with $i$ the negative helicity gluon or photon.  The term associated
with each photon is simply the eikonal factor written in
spinor-helicity notation,
\begin{equation}
{\spa2.1 \over \spa2.j\spa{j}.1} =  \sqrt{2}\left[{\pol^+_j\cdot
        k_1\over s_{1j}} - {\pol^+_j\cdot k_2 \over s_{j2}}\right].
\end{equation}

\subsection{Four-quark amplitudes}
\label{sec:twothree}

For four quarks and $(n-4)$ gluons the color decomposition of the
tree-level amplitude is~\cite{mpReview}
\begin{eqnarray}
\lefteqn{ {\cal A}_n^\tree(\qb_1,q_2,\overline{Q}_3,Q_4,g_5,\ldots,g_n)
= g^{n-2} \sum_{k=0}^{n-4} \sum_{\sigma\in S_{k}}\,
\sum_{\rho\in S_{l}} }
\label{FourQuarkGluonDecomp}\\
&\times
& \Biggl[ (T^{\sigma_1} \dots T^{\sigma_k})_{i_4}^{\ib_1}\
  (T^{\rho_1} \dots T^{\rho_l})_{i_2}^{\ib_3}\
    A_n^\tree(\qb_1,q_2;\overline{Q}_3,Q_4;g_{\sigma_1},\dots,
  g_{\sigma_k};g_{\rho_1},\dots,g_{\rho_l}) \nonumber\\
&-&
  {1\over N_c}\, (T^{\sigma_1} \dots T^{\sigma_k})_{i_2}^{\ib_1}\
  (T^{\rho_1} \dots T^{\rho_l})_{i_4}^{\ib_3}\
    B_n^\tree(\qb_1,q_2;\overline{Q}_3,Q_4;g_{\sigma_1},\dots,
  g_{\sigma_k};g_{\rho_1},\dots,
  g_{\rho_l}) \Biggr]\, ,\nonumber
\end{eqnarray}
with $k+l=n-4$. The sums are over the partitions of $(n-4)$
gluons between the two quark lines, and over the permutations of the
gluons within each partition. For $k=0$ or $l=0$, the color strings
reduce to Kronecker delta's.  For identical quarks, we must subtract
from this equation the same term with the quarks (but not the
anti-quarks) exchanged $(q_2\leftrightarrow Q_4)$.

For maximally-helicity-violating configurations, $(-,-,+,\dots,+)$,
where because of helicity conservation along the fermion lines all
of the gluons have the same helicity, the $A_n^\tree$ and 
$B_n^\tree$ sub-amplitudes factorize into distinct contributions for
the two quark antenn\ae~\cite{Color,mpReview},
\begin{eqnarray}
\lefteqn{ A_n^\tree(\qb_1,q_2;\Qb_3,Q_4;g_{\sigma_1},\dots,
g_{\sigma_k};g_{\rho_1},\dots,g_{\rho_l}) = }
\label{FourQuarkMHV}\\ & &
i{f(\lambda_{q_2},\lambda_{Q_4}) \over \spa{\qb_1}.{q_2}
\spa{\Qb_3}.{Q_4} }\,
{ \spa{q_2}.{\Qb_3} \over \spa{q_2}.{g_{\sigma_1}} \cdots
\spa{g_{\sigma_k}}.{\Qb_3} }\, { \spa{Q_4}.{\qb_1} \over
\spa{Q_4}.{g_{\rho_1}} \cdots \spa{g_{\rho_l}}.{\qb_1} } \nonumber\\
\lefteqn{ B_n^\tree(\qb_1,q_2;\Qb_3,Q_4;g_{\sigma_1},\dots,
g_{\sigma_k};g_{\rho_1},\dots,g_{\rho_l}) = } \nonumber\\ & &
i{f(\lambda_{q_2},\lambda_{Q_4}) \over \spa{\qb_1}.{q_2}
\spa{\Qb_3}.{Q_4} }\,
{ \spa{q_2}.{\qb_1} \over \spa{q_2}.{g_{\sigma_1}} \cdots
\spa{g_{\sigma_k}}.{\qb_1} }\, { \spa{Q_4}.{\Qb_3} \over
\spa{Q_4}.{g_{\rho_1}} \cdots \spa{g_{\rho_l}}.{\Qb_3} }\, ,\nonumber
\end{eqnarray}
with
\begin{eqnarray}
f(+,+) = - \spa{\qb_1}.{\Qb_3}^2 \quad & \quad
f(+,-) = \spa{\qb_1}.{Q_4}^2 \label{pref}\\
f(-,+) = \spa{q_2}.{\Qb_3}^2 \quad & \quad
f(-,-) = - \spa{q_2}.{Q_4}^2\, .\nonumber
\end{eqnarray}
This factorization does not occur for more complicated helicity
configurations.

When we convert a gluon into a photon we attach
the photon everywhere along the two quark lines, i.e. sum over all of the
permutations of the photon index in the global color ordering, and
we replace the color charge $g$ with the appropriate quark electric charge
$\sqrt{2}\, e Q_{q_i}$. The ensuing color decomposition of the amplitude
for the emission of $m$ photons and $r$ gluons, with $m+r=n-4$, is
\begin{eqnarray}
\lefteqn{ {\cal A}_n^\tree(\qb_1,q_2;\Qb_3,Q_4;g_5, \dots,g_{r+4};
\gamma_{r+5}, \dots,\gamma_n) = g^{r+2}\, (\sqrt{2} e)^m\,
\sum_{k=0}^{r} \sum_{\sigma\in S_{k}}\sum_{\rho\in S_{l}} }
\label{FourQuarkGluonPhot}\\ &\times&
\Biggl[ (T^{\sigma_1} \dots T^{\sigma_k})_{i_4}^{\ib_1}\
  (T^{\rho_1} \dots T^{\rho_l})_{i_2}^{\ib_3} \nonumber\\ && \qquad
\times\;  A_n^\tree(\qb_1,q_2;\Qb_3,Q_4;g_{\sigma_1},\dots,
g_{\sigma_k};g_{\rho_1},\dots,g_{\rho_l}; \gamma_{r+5}, \dots,\gamma_n))
\nonumber\\ &-&
{1\over N_c}\, (T^{\sigma_1} \dots T^{\sigma_k})_{i_2}^{\ib_1}\
  (T^{\rho_1} \dots T^{\rho_l})_{i_4}^{\ib_3}\nonumber\\ && \qquad
\times\;  B_n^\tree(\qb_1,q_2;\Qb_3,Q_4;g_{\sigma_1},\dots,
g_{\sigma_k};g_{\rho_1},\dots,
g_{\rho_l}; \gamma_{r+5}, \dots,\gamma_n)) \Biggr]\, ,\nonumber
\end{eqnarray}
with $k+l=r$. The dependence on the
quark electric charges $Q_{p}$ is contained in the sub-amplitudes
$A_n^\tree$ and $B_n^\tree$. For maximally-helicity-violating
configurations, $(-,-,+,\dots,+)$, with  all of the
gluons and photons having positive helicity, 
the sub-amplitudes are~\cite{Color,mpReview},
\begin{eqnarray}
\lefteqn{ A_n^\tree(\qb_1,q_2;\Qb_3,Q_4;g_{\sigma_1},\dots,
g_{\sigma_k};g_{\rho_1},\dots,g_{\rho_l};\gamma_{r+5}, \dots,\gamma_n) = }
\label{FourQuarkGluonPhotMHV}\\ & &
\prod_{j=r+5}^{n} \left(
{ Q_{q}\,\br{q_2}{\qb_1} \over \br{q_2}{\gamma_j}
\br{\gamma_j}{\qb_1} } +  { Q_{Q}\,\br{Q_4}{\Qb_3} \over \br{Q_4}{\gamma_j}
\br{\gamma_j}{\Qb_3} }\right)\,
 A_{n-m}^\tree(\qb_1,q_2;\Qb_3,Q_4;g_{\sigma_1},\dots,
g_{\sigma_k};g_{\rho_1},\dots,g_{\rho_l})\nonumber
\end{eqnarray}
and the same for $B_n^\tree$.

\section{One-Photon Amplitudes}
\label{sec:onephotamps}

One-photon amplitudes with five and six external legs can be used for
NLO calculations of direct photo-production of three jets at
electron-proton colliders \cite{zeus} or photon + two-jet production
at hadron colliders. They also contribute to NNLO calculations of the
same processes with one less jet in the final state.

The tree-level amplitudes with five external legs involve only
maximally-helicity-violating configurations and are easily 
obtained from the formul\ae\ of the previous section.

The one-photon tree-level amplitudes with six external legs are
\begin{itemize}
\item[$a)$] two-quark, three-gluon, one-photon amplitudes;
\item[$b)$] four-quark, one-gluon, one-photon amplitudes.
\end{itemize}

\subsection{Two-quark, three-gluon, one-photon amplitudes}
\label{sec:2q3ggamma}

The color decomposition of the two-quark, three-gluon, one-photon
amplitudes is given in \eqn{TwoQuarkGluonPhot}, with $r=3$ and $m=1$.
For maximally-helicity-violating configurations, the sub-amplitudes are
given in eq.~(\ref{mhvphotamp}).  Making use of
discrete symmetries, there are three distinct helicity configurations
of type $(-,-,-,+,+,+)$, which are listed in \tab{2q3g1gamma}.
\begin{table}[ht]
\begin{center}
\begin{tabular}{|c|cc|cccc|}
\hline
& $\qb_5 $ & $q_6$ & $g_1$ & $g_2$ & $g_3$ & $\gamma_4$ \\
\hline
I   & $+$ & $-$ & $-$ & $-$ & $+$ & $+$ \\
II  & $+$ & $-$ & $+$ & $-$ & $-$ & $+$ \\
III & $+$ & $-$ & $-$ & $+$ & $-$ & $+$ \\
\hline
\end{tabular}
\end{center}
\caption[]{
\label{2q3g1gamma}
\small The distinct helicity configurations of type $(-,-,-,+,+,+)$,
for the two-quark, three-gluon, one-photon amplitudes.
\smallskip}
\end{table}

For these configurations, the sub-amplitudes can be written as 
\begin{eqnarray}
\label{eq:2q3g1gamma1}
-iA_6^{\rm tree}(\qb_5^+,q_6^-,g_1^-,g_2^-,g_3^+,\gamma_4^+)&=&
  {\sapp{3}{1}{2}{5} \sapp{4}{3}{5}{2} \t{1}{2}{6} \over
    \sq{1}{2} \br{4}{5} \sq{1}{6} \br{3}{5} \s{2}{3} \s{4}{6} }
 + {\sapp{3}{2}{5}{1} \sapp{4}{3}{5}{2} ^2\over
    \br{3}{5} \sq{1}{6} \s{2}{3} \s{4}{6} \t{1}{4}{6}}\nonumber\\
&-& {\br{1}{2} \sq{4}{5}\sapp{3}{1}{2}{6}\sapp{3}{1}{2}{5}\over
    \sq{1}{2} \br{4}{5} \s{2}{3} \s{4}{6} \t{1}{2}{3}}\\[5pt]
\label{eq:2q3g1gamma2}
-iA_6^{\rm tree}(\qb_5^+,q_6^-,g_1^+,g_2^-,g_3^-,\gamma_4^+)&=&
   {\br{2}{3} \sapp{1}{2}{3}{6}^2 \over
    \sq{2}{3} \br{5}{4} \br{4}{6} \s{1}{2} \t{1}{2}{3} }
  + {\br{2}{6} \sapp{1}{2}{3}{6} \sapp{1}{2}{6}{3} \t{1}{4}{6}
   \over \sq{2}{3} \br{5}{4} \br{4}{6} \s{1}{2} \s{1}{6} \s{3}{5} }\nonumber\\
&+& {\br{2}{3} \br{2}{6} \sq{4}{1} \sapp{5}{1}{4}{6}
   \over \sq{2}{3} \br{1}{2} \br{4}{6} \s{1}{6} \s{3}{5} }
   + {\br{2}{6} \sapp{1}{2}{6}{3}^2 \sapp{4}{3}{5}{2} \over
    \br{5}{4} \s{1}{2} \s{1}{6} \s{3}{5} \t{1}{2}{6} }\nonumber\\
\\
\label{eq:2q3g1gamma3}
-iA_6^{\rm tree}(\qb_5^+,q_6^-,g_1^-,g_2^+,g_3^-,\gamma_4^+)&=&
  - {\br{1}{3}^2 \sapp{2}{1}{3}{6}^2 \over
      \br{5}{4} \br{4}{6} \s{1}{2} \s{2}{3} \t{1}{2}{3} }
  + {\sq{2}{6} \sq{4}{5} \br{6}{1} \br{5}{3} \sapp{2}{4}{5}{3} \over
      \sq{1}{6} \br{5}{4} \s{1}{2} \s{3}{5} \t{1}{2}{6} }\nonumber\\
&+& {\sq{2}{5}^2 \br{6}{1} \sapp{4}{2}{5}{3} \over
      \sq{1}{6} \sq{5}{3} \br{4}{6} \s{2}{3} \t{1}{4}{6} }
  + {\br{1}{6} \sq{2}{5} \sapp{2}{1}{3}{5} \sapp{2}{1}{6}{3} \over
      \sq{1}{6} \sq{2}{3} \br{5}{4} \br{4}{6} \s{1}{2} \s{3}{5} }\nonumber\\
&+& {\br{1}{3} \sapp{2}{1}{3}{6} \sapp{2}{1}{6}{3} \over
      \sq{1}{6} \br{5}{4} \br{4}{6} \s{1}{2} \s{2}{3} }
  + {\br{1}{6} \sq{2}{5} \br{1}{3} \sapp{4}{2}{5}{3} \over
      \sq{1}{6} \br{4}{6} \br{2}{1} \s{2}{3} \s{3}{5} }\nonumber\\
\end{eqnarray}

These amplitudes have been computed directly from the color-ordered
Feynman diagrams.  Equivalent but less compact expressions could be
obtained by summing over permutations of the two-quark four-gluon
amplitudes,
\begin{eqnarray}
&&\hspace{-2cm} A_6^{\rm tree}(\qb_5,q_6,g_1,g_2,g_3,\gamma_4)=
 A_6^{\rm tree}(\qb_5,q_6,g_1,g_2,g_3,g_4)  +
 A_6^{\rm tree}(\qb_5,q_6,g_1,g_2,g_4,g_3)
\nonumber\\&& \hspace{2cm} A_6^{\rm tree}
(\qb_5,q_6,g_1,g_4,g_2,g_3) + A_6^{\rm tree}
(\qb_5,q_6,g_4,g_1,g_2,g_3) \,.
\end{eqnarray}
An alternative method is to use supersymmetric Ward
identities to obtain the result from the known two-quark four-gluon
and six-gluon amplitudes (see Appendix~\ref{app:2q3g1gamma}).

\subsection{Four-quark, one-gluon, one-photon amplitudes}
\label{sec:fourphotamps}

The color decomposition of the four-quark, one-gluon, one-photon
amplitudes is given in eq.~(\ref{FourQuarkGluonPhot}), with $m=r=1$.
For each helicity configuration there are four independent color 
structures.  For the configurations of type $(-,-,+,+,+,+)$, the
amplitudes are given in eq.~(\ref{FourQuarkGluonPhotMHV}),
with $m=1$ and $(k,l) = (1,0)$ and $(0,1)$.

\Tab{4q1g1gamma} lists the two distinct helicity configurations of type
$(-,-,-,+,+,+)$.  
%
\begin{table}[ht]
\begin{center}
\begin{tabular}{|c|cccc|cc|}
\hline
& $\qb_1$ & $q_2$ & $\Qb_3$ & $Q_4$ & $g_5$ & $\gamma_6$ \\
\hline
I   &  +  & $-$ & $+$ & $-$ & $-$ & $+$ \\
II  &  +  & $-$ & $-$ & $+$ & $-$ & $+$ \\
\hline
\end{tabular}
\end{center}
\caption[]{
\label{4q1g1gamma}
\small The distinct helicity configurations of type $(-,-,-,+,+,+)$,
for the four-quark, one-gluon, one-photon amplitudes.
\smallskip}
\end{table}
%
The color-ordered sub-amplitudes for these configurations are
\begin{eqnarray}
&&\hspace*{-2cm}
\nonumber\\&&\hspace*{-2cm}
A_6^{\rm tree}(\qb_1^+,q_2^-;\Qb_3^+,Q_4^-;\emptyset;g_5^-;\gamma_6^+)=
Q_q f_1(1,2,3,4;5;6)+ Q_Q f_2(3,4,1,2;5;6)
\nonumber\\[2pt]&& \hspace*{-2cm}
A_6^{\rm tree}(\qb_1^+,q_2^-;\Qb_3^+,Q_4^-;g_5^-;\emptyset;\gamma_6^+)=
Q_q f_2(1,2,3,4;5;6)+ Q_Q f_1(3,4,1,2;5;6)
\nonumber\\[-8pt]\label{4q1g1gammaI} \\[-8pt]&& \hspace*{-2cm}
B_6^{\rm tree}(\qb_1^+,q_2^-;\Qb_3^+,Q_4^-;g_5^-;\emptyset;\gamma_6^+) =
Q_q g_1(1,2,3,4;5;6)+ Q_Q g_2(3,4,1,2;5;6)
\nonumber\\[2pt]&& \hspace*{-2cm}
B_6^{\rm tree}(\qb_1^+,q_2^-;\Qb_3^+,Q_4^-;\emptyset;g_5^-;\gamma_6^+) =
Q_q g_2(1,2,3,4;5;6)+ Q_Q g_1(3,4,1,2;5;6)
\nonumber
\end{eqnarray}
\begin{eqnarray}
&&\hspace*{-2cm}
\nonumber\\&&\hspace*{-2cm}
A_6^{\rm tree}(\qb_1^+,q_2^-;\Qb_3^-,Q_4^+;\emptyset;g_5^-;\gamma_6^+)=
\phantom{-}Q_q f_3(1,2,3,4;5;6)+ Q_Q f_3(4,3,2,1;5;6) 
\nonumber\\[2pt]&& \hspace*{-2cm}
A_6^{\rm tree}(\qb_1^+,q_2^-;\Qb_3^-,Q_4^+;g_5^-;\emptyset;\gamma_6^+)=
\phantom{-}Q_q f_4(1,2,3,4;5;6)+ Q_Q f_4(4,3,2,1;5;6) 
\nonumber\\[-8pt]\label{4q1g1gammaII} \\[-8pt]&& \hspace*{-2cm}
B_6^{\rm tree}(\qb_1^+,q_2^-;\Qb_3^-,Q_4^+;g_5^-;\emptyset;\gamma_6^+) =
\phantom{-}Q_q g_1(1,2,4,3;5;6)- Q_Q g_2(4,3,1,2;5;6)
\nonumber\\[2pt]&& \hspace*{-2cm}
B_6^{\rm tree}(\qb_1^+,q_2^-;\Qb_3^-,Q_4^+;\emptyset;g_5^-;\gamma_6^+) =
-Q_q g_2(1,2,4,3;5;6)+ Q_Q g_1(4,3,1,2;5;6) 
\nonumber
\end{eqnarray}
where
\begin{eqnarray}
-if_1(1,2,3,4;5;6) &=&
    {\sapp{3}{1}{6}{2}\sapp{3}{1}{6}{4} \over
      \br{1}{6}\br{6}{2}\sq{3}{5}\sq{5}{2}\s{3}{4}}\nonumber\\
 &&+{\br{5}{2}\sq{3}{1}\sapp{6}{2}{5}{4} \over
      \sq{5}{2}\br{6}{2}\s{3}{4}\t{1}{3}{4}}
   +{\br{4}{5}\sapp{3}{1}{6}{2}^2 \over
      \br{1}{6}\br{6}{2}\sq{3}{5}\s{3}{4}\t{1}{2}{6}}\label{4q1g1gammaf1}\\[7pt] 
-if_2(1,2,3,4;5;6) &=&
    {\sapp{1}{4}{5}{2}\sapp{3}{1}{6}{2} \over
      \br{1}{6}\br{6}{2}\sq{1}{5}\sq{5}{4}\s{3}{4}}\nonumber\\
 &&+{\sq{1}{6}\br{2}{4}\sapp{3}{1}{6}{5} \over
      \br{1}{6}\sq{1}{5}\s{3}{4}\t{1}{5}{6}}
   +{\br{5}{3}\sapp{3}{1}{6}{2}^2 \over
      \br{1}{6}\br{6}{2}\sq{5}{4}\s{3}{4}\t{1}{2}{6}}\label{4q1g1gammaf2}\\
-ig_1(1,2,3,4;5;6) &=&
    {\sapp{1}{3}{6}{4}\sapp{3}{1}{6}{2} \over
      \br{1}{6}\br{6}{2}\sq{1}{5}\sq{5}{2}\s{3}{4}}\nonumber\\
 &&+{\sq{1}{6}\br{2}{4}\sapp{3}{1}{6}{5} \over
      \br{1}{6}\sq{1}{5}\s{3}{4}\t{1}{5}{6}}
   +{\br{5}{2}\sq{3}{1}\sapp{6}{2}{5}{4} \over
      \sq{5}{2}\br{6}{2}\s{3}{4}\t{1}{3}{4}}\label{4q1g1gammag1}\\[7pt]
-ig_2(1,2,3,4;5;6) &=&
   -{\sapp{3}{1}{6}{2}^2 \over
      \br{1}{6}\br{6}{2}\sq{3}{5}\sq{5}{4}\t{1}{2}{6}}
\label{4q1g1gammag2}\\[7pt]
-if_3(1,2,3,4;5;6) &=&
   -{\t{1}{4}{6}\sapp{4}{1}{6}{2} \over
      \br{1}{6}\br{6}{2}\sq{3}{5}\sq{5}{2}\s{3}{4}}\nonumber\\
 &&+{\br{5}{4}\sapp{4}{1}{6}{2}^2 \over
      \br{1}{6}\br{6}{2}\sq{3}{5}\s{3}{4}\t{1}{2}{6}}
   +{\br{5}{2}\sq{4}{1}\sapp{6}{2}{5}{3} \over
      \sq{5}{2}\br{6}{2}\s{3}{4}\t{1}{3}{4}}\label{4q1g1gammaf3}\\[7pt]
-if_4(1,2,3,4;5;6) &=&
    {\sq{4}{1}\br{2}{3}\sapp{4}{1}{6}{2} \over
      \br{1}{6}\br{6}{2}\sq{1}{5}\sq{5}{4}\s{3}{4}}\nonumber\\
 &&+{\sq{1}{6}\br{2}{3}\sapp{4}{1}{6}{5} \over
      \br{1}{6}\sq{1}{5}\s{3}{4}\t{1}{5}{6}}
   +{\br{3}{5}\sapp{4}{1}{6}{2}^2 \over
      \br{1}{6}\br{6}{2}\sq{5}{4}\s{3}{4}\t{1}{2}{6}}\label{4q1g1gammaf4}
\end{eqnarray}

All other configurations
are related to these by parity inversion and charge conjugation\footnote{
When photons are present in the sub-amplitudes, we apply charge conjugation
on the kinematic part only, without flipping the sign of the quark
electric charge.}. In
performing charge conjugation, one must exchange each quark with its
anti-quark ($ \qb_i \leftrightarrow q_i$ for $i=1,2$).
For example, by applying charge conjugation and parity inversion on
eq.(\ref{4q1g1gammaI}), we obtain
\begin{eqnarray}
&&\hspace*{-2cm}
A_6^{\rm tree}(\qb_1^-,q_2^+;\Qb_3^-,Q_4^+;\emptyset;g_5^-;\gamma_6^+)=
Q_q f_2(2,1,4,3;5;6)+ Q_Q f_1(4,3,2,1;5;6)
\nonumber\\[2pt]&& \hspace*{-2cm}
A_6^{\rm tree}(\qb_1^-,q_2^+;\Qb_3^-,Q_4^+;g_5^-;\emptyset;\gamma_6^+)=
Q_q f_1(2,1,4,3;5;6)+ Q_Q f_2(4,3,2,1;5;6)
\nonumber\\[-8pt]\label{4q1g1gammaIV} \\[-8pt]&& \hspace*{-2cm}
B_6^{\rm tree}(\qb_1^-,q_2^+;\Qb_3^-,Q_4^+;g_5^-;\emptyset;\gamma_6^+) =
Q_q g_1(2,1,4,3;5;6)+ Q_Q g_2(4,3,2,1;5;6)
\nonumber\\[2pt]&& \hspace*{-2cm}
B_6^{\rm tree}(\qb_1^-,q_2^+;\Qb_3^-,Q_4^+;\emptyset;g_5^-;\gamma_6^+) =
Q_q g_2(2,1,4,3;5;6)+ Q_Q g_1(4,3,2,1;5;6)\, .\nonumber
\end{eqnarray}

\section{Two-Photon Amplitudes}
\label{sec:twophotamps}

The amplitudes with two photons can be used for NLO two-photons +
one-jet production at hadron colliders, for NLO photon + two-jet
production in direct photo-production at electron-proton colliders, or
for NLO three-jet production in gamma-gamma scattering at some future
$e^+e^-$ collider. They also contribute to NNLO calculations of the
same processes with one less jet in the final state.

These amplitudes are of particular importance to the evaluation of
the irreducible backgrounds to low-mass Higgs searches at LHC;
they contribute to the NLO evaluation of the background to Higgs + 
one-jet production and to the NNLO evaluation of the background to 
inclusive Higgs production.

The tree-level two-photon amplitudes with six external legs are
\begin{itemize}
\item[$a)$] two-quark, two-gluon, two-photon amplitudes; 
\item[$b)$] four-quark, two-photon amplitudes.
\end{itemize}

\subsection{Two-quark, two-gluon, two-photon amplitudes}
\label{sec:2q2g2gamma}

The color decomposition of the two-quark, two-gluon, two-photon amplitudes
is given in eq.~(\ref{TwoQuarkGluonPhot}). For the
helicity configurations of type $(-,-,+,+,+,+)$, the sub-amplitudes are
given in eq.~(\ref{mhvphotamp}), with $r=m=2$.
There are three distinct configurations of type $(-,-,-,+,+,+)$,
which are listed in \tab{2q2g2gamma}. The other helicity
configurations are obtained from these by particle relabeling, parity
inversion and/or charge conjugation. 
\begin{table}[ht]
\begin{center}
\begin{tabular}{|c|cc|cccc|}
\hline
& $\qb_5 $ & $q_6$ & $g_1$ & $g_2$ & $\gamma_3$ & $\gamma_4$ \\
\hline
I   & $+$ & $-$ & $-$ & $-$ & $+$ & $+$ \\
II & $+$ & $-$ & $+$ & $-$ & $-$ & $+$ \\
III  & $+$ & $-$ & $-$ & $+$ & $-$ & $+$ \\
\hline
\end{tabular}
\end{center}
\caption[]{
\label{2q2g2gamma}
\small The distinct helicity configurations of type $(-,-,-,+,+,+)$,
for the two-quark, two-gluon, two-photon amplitudes.
\smallskip}
\end{table}
\begin{eqnarray}
\label{eq:2q2g2gamma1}
-iA_6^{\rm tree}(\qb_5^+,q_6^-,g_1^-,g_2^-,\gamma_3^+,\gamma_4^+)&=&
     {\sapp{5}{1}{2}{6} \br{5}{6} \t{1}{2}{6}
      \over \sq{5}{2} \sq{1}{6} \sq{2}{1}
            \br{6}{3} \br{6}{4} \br{5}{3} \br{5}{4}}\nonumber\\
   &-& {\br{1}{6} \sq{4}{5} \sapp{3}{1}{6}{2}
      \over \sq{1}{6} \br{4}{5} \sq{5}{2} \br{6}{3} \t{1}{3}{6}}
    -  {\br{1}{6} \sq{3}{5} \sapp{4}{1}{6}{2}
      \over \sq{1}{6} \br{3}{5} \sq{5}{2} \br{6}{4} \t{1}{4}{6}}\nonumber\\
\\
\label{eq:2q2g2gamma2}
-iA_6^{\rm tree}(\qb_5^+,q_6^-,g_1^+,g_2^-,\gamma_3^-,\gamma_4^+)&=&
       {{\sq{4}{5} \br{2}{6}}^2 \sapp{1}{2}{6}{3}
       \over \br{4}{5} \sq{5}{3} \br{1}{6} \s{1}{2}\t{1}{2}{6}}
    +  {\br{3}{6} \sq{5}{4} \sapp{1}{3}{6}{2}
       \over  \sq{3}{6} \br{5}{4} \sq{5}{2} \br{1}{6}\t{1}{3}{6}}\nonumber\\
   &+& {\br{3}{6} {\sq{1}{5}}^2 \sapp{4}{3}{6}{2}
       \over  \sq{3}{6} \sq{2}{5} \br{6}{4} \s{1}{2}\t{1}{2}{5}}
    +  {\sq{5}{1} \br{6}{2} \sapp{5}{1}{4}{6}\t{1}{4}{5}
       \over \sq{3}{6} \sq{5}{2} \sq{5}{3}
           \br{6}{1} \br{6}{4} \br{4}{5} \s{1}{2}}\nonumber\\
   &+& {\sapp{1}{3}{6}{2}\left( \sq{5}{1} \br{1}{6} \sapp{5}{2}{3}{6}
                     + \sq{5}{4} \br{4}{6} \sq{5}{2} \br{2}{6}\right)
       \over \sq{3}{6} \sq{5}{2} \sq{5}{3}
           \br{6}{1} \br{6}{4} \br{4}{5} \s{1}{2}}\nonumber\\
\\
\label{eq:2q2g2gamma3}
-iA_6^{\rm tree}(\qb_5^+,q_6^-,g_1^-,g_2^+,\gamma_3^-,\gamma_4^+)&=&
       {\br{1}{6} \sq{2}{5} \sapp{4}{2}{5}{3}
       \over \sq{1}{6} \br{2}{5} \sq{5}{3} \br{6}{4} \t{1}{4}{6}}
    +  {\br{1}{6} \sq{4}{5} \sq{6}{2} \sapp{2}{4}{5}{3}
       \over \sq{1}{6} \br{4}{5} \sq{5}{3} \s{1}{2} \t{1}{2}{6}}\nonumber\\
   &+& {\br{3}{6} \sq{2}{5} \br{5}{1} \sapp{4}{2}{5}{1}
       \over \sq{3}{6} \br{2}{5} \br{6}{4} \s{1}{2} \t{3}{4}{6}}
    +  {\br{1}{6} \sq{2}{5} \sapp{2}{4}{5}{3}
        \over \sq{1}{6} \sq{3}{5} \br{5}{4} \br{4}{6} \s{1}{2}}\nonumber\\
   &+& {\t{1}{3}{6}\left(
        \br{5}{1} \sq{2}{6} \sq{5}{3} \br{3}{6}
      + \sq{1}{2} \sq{6}{5} \br{1}{6} \br{5}{1}\right)
        \over \sq{1}{6} \sq{5}{3} \sq{3}{6} \br{2}{5}
        \br{5}{4} \br{4}{6} \s{1}{2}}\nonumber\\
\end{eqnarray}

These expressions were obtained by direct computation of the
color-ordered Feynman diagrams.  Equivalent representations can easily
be obtained by expressing the amplitudes as linear combinations of the
one photon amplitudes,
\begin{eqnarray}
&&\hspace{-2cm} A_6^{\rm tree}(\qb_5,q_6,g_1,g_2,\gamma_3,\gamma_4)=
 A_6^{\rm tree}(\qb_5,q_6,g_1,g_2,g_3,\gamma_4)  +
\nonumber\\&& \hspace{2cm} A_6^{\rm tree}
(\qb_5,q_6,g_3,g_1,g_2,\gamma_4) + A_6^{\rm tree}
(\qb_5,q_6,g_1,g_3,g_2,\gamma_4) \,.
\end{eqnarray}
Alternatively, they can be obtained from two-quark four-gluon and
six-gluon amplitudes using supersymmetric Ward identities
(see Appendix~\ref{app:2q2g2gamma}).

\subsection{Four-quark, two-photon amplitudes}
\label{sec:4q2gamma}

The color decomposition of the four-quark, two-photon
amplitudes is given in eq.~(\ref{FourQuarkGluonPhot}).
For the helicity configurations of type $(-,-,+,+,+,+)$, the
sub-amplitudes are given in eq.~(\ref{FourQuarkGluonPhotMHV}),
with $m=2$ and $r=0$.  The color strings reduce to Kronecker delta's,
so that the amplitude is written as a single kinematic term
multiplying the color factor
$$\delta_{i_4}^{\ib_1}\delta_{i_2}^{\ib_3}
  - {1\over N}\delta_{i_2}^{\ib_1}\delta_{i_4}^{\ib_3}.$$
There are two distinct helicity configurations of type
$(-,-,-,+,+,+)$ which are listed in \tab{4q2gamma}.  All other
configurations are related to these by parity inversion and charge
conjugation.
\begin{table}[ht]
\begin{center}
\begin{tabular}{|c|cccc|cc|}
\hline
& $\qb_1$ & $q_2$ & $\Qb_3$ & $Q_4$ & $\gamma_5$ & $\gamma_6$ \\
\hline
I   &  +  & $-$ & $+$ & $-$ & $-$ & $+$ \\
II  &  +  & $-$ & $-$ & $+$ & $-$ & $+$ \\
\hline
\end{tabular}
\end{center}
\caption[]{
\label{4q2gamma}
\small The distinct helicity configurations of type $(-,-,-,+,+,+)$,
for the four-quark, two-photon amplitudes.
\smallskip}
\end{table}
The amplitudes can be written in terms of two functions $g_1$ and
$g_2$, which are identical to the $g_{1,2}$ of
\Eqns{4q1g1gammag1}{4q1g1gammag2},
%
%
\begin{eqnarray}
A_6^{\rm tree}(\qb_1^+,q_2^-;\Qb_3^+,Q_4^-;\gamma_5^-;\gamma_6^+)&=&
Q_q^2 g_1(1,2,3,4;5,6)+Q_Q^2 g_1(3,4,1,2;5,6)\nonumber\\
&+&Q_q Q_Q(g_2(1,2,3,4;5,6)+g_2(3,4,1,2;5,6))\nonumber\\
\\
A_6^{\rm tree}(\qb_1^+,q_2^-;\Qb_3^-,Q_4^+;\gamma_5^-;\gamma_6^+)&=&
Q_q^2 g_1(1,2,4,3;5,6)+Q_Q^2 g_1(4,3,1,2;5,6)\nonumber\\
&-&Q_q Q_Q(g_2(1,2,4,3;5,6)+g_2(4,3,1,2;5,6))\, .\nonumber
\end{eqnarray}

\section{Three-Photon Amplitudes}
\label{sec:threephotamps}

The amplitudes with three photons can be used for NLO three-photon
production at hadron colliders, for NLO two-photon inclusive production
and two-photon + one-jet
production in direct photo-production at electron-proton colliders, or
for NLO photon + one- or two-jet inclusive production in 
gamma-gamma collisions.

For NLO calculations we need two-quark, one-gluon, three-photon
tree-level amplitudes.  For the
helicity configurations of type $(-,-,+,+,+,+)$, the sub-amplitudes are
given in eq.~(\ref{mhvphotamp}), with $r=1$ and $m=3$.
There is one distinct helicity configuration of type $(-,-,-,+,+,+)$,
for which the sub-amplitude is
\begin{eqnarray}
&&\hspace*{-2cm}
-iA_6^{\rm tree}(\qb^+_5,q^-_6,g_1^-,\gamma_2^-,\gamma_3^+,\gamma_4^+)=
\nonumber\\
&&\hspace*{-1cm}\frac{s_{5  6} t_{6 1 2}
\sapp{5}{3}{4}{6}}
 {\br{5}{3} \br{3}{ 6} \,
 \br{5}{ 4} \br{4}{ 6} \,
 \sq{5}{ 1} \sq{1}{6} \,
 \sq{5}{ 2} \sq{2}{6}}
+\sum_{\sigma \in S_2, \rho \in S_2}
\frac{
 \sq{5}{\rho_3}\br{6}{\sigma_2} \sap{\rho_4}{5}{\rho_3}{\sigma_1} }
{\br{5}{\rho_3}\sq{6}{\sigma_2} \br{6}{\rho_4} \sq{5}{\sigma_1}
 t_{6  \rho_4  \sigma_2} } \,.
\end{eqnarray}
We note that this is also the sub-amplitude for the two-quark,
four-photon process, since only the color factor differs.  Indeed,
this sub-amplitude is also identical to that obtained for
$e^+e^-\to\gamma\gamma\gamma\gamma$~\cite{calkul}.

\section{One-Loop Amplitudes}
\label{sec:loops}

The results in this section are obtained either in their entirety or
from discussions contained in references~\cite{BDK,AS}.  Our
purpose here is to present all of the results in terms of known
components so that they can be combined with the tree-level results of
the previous sections to produce complete NLO calculations.
When supplemented with higher-order terms in $\epsilon$, they also
contribute to NNLO calculations.

The color decomposition of one-loop amplitudes in QCD is somewhat more
complicated than it is for tree-level amplitudes.  If there are
external gluons, one finds new color structures at one-loop.
Associated with each color structure is a ``partial
amplitude''\cite{BDK}.  For instance, the two-quark, three-gluon
one-loop matrix element is written as
\begin{eqnarray}
\label{eq:QQGGG}
{\cal A}^\oneloop(\qb_1,q_2;g_3,g_4,g_5) &=& g^5\left\{
            \sum_{\sigma\in S_3}N_c
            \left(T^{\sigma_3}T^{\sigma_4}T^{\sigma_5}
            \right)^{\ib_1}_{i_2}A_{5;1}
            (\qb_1,q_2;g_{\sigma_3},g_{\sigma_4},g_{\sigma_5})
            \right.\nonumber \\
            &+& \phantom{g^5\left\{\right.}\sum_{\sigma\in S_3}
            \Tr\left(T^{\sigma_3}\right)
            \left(T^{\sigma_4}T^{\sigma_5}\right)^{\ib_1}_{i_2}
            A_{5;2}(\qb_1,q_2;g_{\sigma_3};g_{\sigma_4},g_{\sigma_5}) \\
            &+& \phantom{g^5}\sum_{\sigma\in S_3/{Z}_2}
            \Tr\left(T^{\sigma_3}T^{\sigma_4}\right)
            \left(T^{\sigma_5}\right)^{\ib_1}_{i_2}
            A_{5;3}(\qb_1,q_2;g_{\sigma_3},
            g_{\sigma_4};g_{\sigma_5})\nonumber \\
            &+& \phantom{g^5}\left.\sum_{\sigma\in S_3/{Z}_3}
            \Tr\left(T^{\sigma_3}T^{\sigma_4}T^{\sigma_5}\right)
            \delta^{\ib_1}_{i_2} A_{5;4}
           (\qb_1,q_2;g_{\sigma_3},g_{\sigma_4},g_{\sigma_5})\right\}\nonumber
\end{eqnarray}
where the sums exclude permutations that leave the color factors
invariant.  For instance, in the third term the trace is unchanged if
elements $3$ and $4$ are interchanged.  Thus $\sigma$ takes values in
the set
\begin{equation}
\label{sumexplain}
S_3/{Z}_2=\{(345), (354), (453)\}.
\end{equation}
Note that the color factor associated with the $A_{5;2}$ term
vanishes, since $\Tr(T^a) = 0$ when $T^a$ is a generator of $SU(N_c)$.
For this reason, $A_{5;2}$ terms are normally dropped.  In the
amplitudes below, we keep the $A_{5;2}$ terms so that one can follow
all of the terms as gluons are converted into photons.


\subsection{Two-quark amplitudes}
The partial amplitudes are most efficiently expressed in terms of a
set of gauge-invariant, color-ordered building blocks called
``primitive amplitudes''~\cite{BDK}.  For the two-quark amplitudes, it
is convenient to construct two primitive amplitudes for each internal
spin state ({\it i.e.\/} spin of the particle going around the loop), 
$A_n^{L,[J]}(\qb_1,g_3,\dots,q_2,\dots,g_n)$ and
$A_n^{R,[J]}(\qb_1,g_3,\dots,q_2,\dots,g_n)$,\ $J=1,{1\over2},0$, with
the $L$ and $R$ primitive amplitudes related by reversing the cyclic
ordering of the external legs:
\begin{equation}
\label{eq:LandR}
A_n^{R,[J]}(\qb_1,g_3,\dots,q_2,\dots,g_n) = (-1)^n
A_n^{L,[J]}(\qb_1,g_n,\dots,q_2,\dots,g_3).
\end{equation}
Spin $1$, of course, refers to gluons in the loop, while spins
${1\over2}$ and $0$ refer to massless ($N_c+\overline{N}_c$)
representation Weyl fermions, and complex scalars respectively.

It turns out that the $R$-type primitives are not needed for
$J={1\over2},0$, so we abbreviate the notation as
\begin{equation}
A_n^{L,[1]}\rightarrow A_n^L \qquad 
A_n^{R,[1]}\rightarrow A_n^R \qquad
A_n^{L,[{1\over2}]}\rightarrow A_n^{[{1\over2}]} \qquad
A_n^{L,[0]}\rightarrow A_n^{[0]}.  
\end{equation}

The $A_{n;1}$ partial amplitudes have the simplest representation in
terms of primitive amplitudes:
\begin{eqnarray}
\label{eq:NIPrim}
A_{n;1}(\qb_1,q_2,g_3,\dots,g_n) &=&
A^{L}_n(\qb_1,q_2,g_3,\dots,g_n) - {1\over
N_c^2}A^{R}_n(\qb_1,q_2,g_3,\dots,g_n)\\
&&+ {n_f\over N_c}A^{[{1\over2}]}_n(\qb_1,q_2,g_3,\dots,g_n) +{n_s\over
N_c}A^{[0]}_n(\qb_1,q_2,g_3,\dots,g_n),\nonumber 
\end{eqnarray}
where $n_f$ is the number of fermions and $n_s$ is the number of
scalars.  In these partial amplitudes, only those primitive amplitudes
with the same ordering of legs as the color factor  contribute.

The partial amplitudes $A_{n;c>1}$ are expressed as sums over
primitive amplitudes with various permutations of the external legs.
For the two-quark amplitudes~\cite{BDK},
\begin{eqnarray}
\label{eq:qqgludecup}
A_{n;c}(\qb_1,q_2;g_3,\dots,g_{c+1};g_{c+2},\dots,g_n)&&\nonumber\\
 &&\hspace{-5cm}= (-1)^{c-1}\sum_{\sigma\in {\rm COP}\{\alpha\}\{\beta\}}
 \left[A_n^{L}(\sigma)\vphantom{{n_f\over N_c}}\right.
\left.-{n_f\over N_c}A_n^{R,[{1\over2}]}(\sigma) - {n_s\over
 N_c}A_n^{R,[0]}(\sigma)\right],
\end{eqnarray}
where $\{\alpha\}\equiv\{g_{c+1},g_c,\dots,g_3\}$ and
$\{\beta\}\equiv\{\qb_1,q_2,g_{c+2},g_{c+3},\dots,g_n\}$, and
${\rm COP}\{\alpha\}\{\beta\}$
is the set of permutations of $\{\qb_1,q_2,\dots,g_n\}$, with the
position of $\qb_1$ held fixed that preserve the cyclic ordering of
the $\alpha_i$ within $\{\alpha\}$ and of the $\beta_j$ within
$\{\beta\}$ while allowing for all possible relative orderings of the
$\alpha_i$ with respect to the $\beta_j$.  Expanding this expression,
and making use of symmetry relations involving
$A^{[{1\over2},0]}$~\cite{BDK}, yields
\begin{eqnarray}
\label{eq:qparts}
A_{5;2}(\qb_1,q_2;g_3;g_4,g_5) &=&
   - A^{L}_5(\qb_1,q_2,g_4,g_5,g_3) 
   - A^{L}_5(\qb_1,q_2,g_4,g_3,g_5)\nonumber\\
&& - A^{L}_5(\qb_1,q_2,g_3,g_4,g_5)
   - A^{L}_5(\qb_1,g_3,q_2,g_4,g_5)\\[10pt]
A_{5;3}(\qb_1,q_2;g_4,g_5;g_3) &=&
   \sum_{\sigma\in S_3}A^{L}_5(\qb_1,q_2,g_{\sigma_3},
   g_{\sigma_4},g_{\sigma_5})
   -\sum_{\sigma\in {Z}_2}A^{R}_5(\qb_1,g_3,q_2,
   g_{\sigma_4},g_{\sigma_5})\nonumber\\
&+&\sum_{\sigma\in {Z}_2}A^{L}_5(\qb_1,g_4,q_2,
   g_{\sigma_3},g_{\sigma_5})+ \sum_{\sigma\in {Z}_2}
   A^{L}_5(\qb_1,g_5,q_2,g_{\sigma_3},g_{\sigma_4})\nonumber\\
\\
A_{5;4}(\qb_1,q_2;g_3,g_4,g_5) &=&
   \sum_{\sigma\in {Z}_3}\left\{\vphantom{A^{[{1\over2}]}_5}
   A^{R}_5(\qb_1,q_2,g_{\sigma_3},g_{\sigma_4},g_{\sigma_5})
 + A^{R}_5(\qb_1,g_{\sigma_3},q_2,g_{\sigma_4},g_{\sigma_5})
 \right.\nonumber\\
&&\phantom{{n_f\over N_c}}
  - A^{L}_5(\qb_1,q_2,g_{\sigma_3},g_{\sigma_5},g_{\sigma_4})
  - A^{L}_5(\qb_1,g_{\sigma_3},q_2,g_{\sigma_5},g_{\sigma_4})\nonumber\\
&&-{n_f\over N_c}\left[
   A^{[{1\over2}]}_5(\qb_1,q_2,g_{\sigma_3},g_{\sigma_4},g_{\sigma_5})
 + A^{[{1\over2}]}_5(\qb_1,g_{\sigma_3},q_2,g_{\sigma_4},g_{\sigma_5})
 \right]\nonumber\\
&&-{n_s\over N_c}\left.\left[
   A^{[0]}_5(\qb_1,q_2,g_{\sigma_3},g_{\sigma_4},g_{\sigma_5})
 + A^{[0]}_5(\qb_1,g_{\sigma_3},q_2,g_{\sigma_4},g_{\sigma_5})
 \right]\vphantom{A^{[{1\over2}]}_5}\right\}.\nonumber\\
\end{eqnarray}

The explicit expressions for the primitive amplitudes are given in
reference~\cite{BDK}.  For reasons of computational simplicity, the
results presented there are in terms of a different linear combination
of primitive amplitudes, labeled $A^L$, $A^{\rm SUSY}$, $A^f$ and $A^s$, which
are related to those used here by
\begin{eqnarray}
A^{L}(\sigma)&=& A^L(\sigma)\nonumber\\
A^{R}(\sigma)&=& A^{\rm SUSY}(\sigma) - A^L(\sigma) + A^f(\sigma) +
                     A^s(\sigma)\nonumber\\ 
A^{[{1\over2}]}(\sigma)&=&-A^f(\sigma) - A^s(\sigma)\\
A^{[0]}(\sigma)&=& A^s(\sigma)\, .\nonumber
\end{eqnarray}
In all generality the expression for $A^{R}(\sigma)$ should have a
$- A^{R,[{1\over 2}]}(\sigma)$ term on the right-hand side. 
For all orderings that appear below this term vanishes.

\subsection{One-loop corrections to ${\qb}qgg\gamma$
amplitudes}
\label{sec:qqgggamma}
This process has been discussed thoroughly by Signer~\cite{AS}.
We present the results directly in terms of the $\qb qggg$ primitive
amplitudes of Bern, Dixon and Kosower~\cite{BDK}.  In order to convert
gluons into photons, we follow the prescription in
reference~\cite{BDK}\footnote{Note that on this topic, the e-print
supersedes the journal reference.}.  This is essentially the same
procedure outlined in \sec{sec:twotwo}, but for a few subtleties that
arise at one loop.  One such subtlety is to note that when there are
fermion loops, the photon can couple to either the external quark line
(external coupling) or to the loop fermion (internal coupling).  For
external coupling, one finds a factor of $Q_qn_f$, the charge of the
external quark times the number of fermion flavors contributing loops.
For internal coupling, one finds a factor of
$\displaystyle\sum_{i=1}^{n_f}Q_{q_i}$, the sum of charges of the
different fermion flavors appearing in the loop, which we abbreviate
as $\Tr\left[Q_{f}\right]$.  A second subtlety is related and involves
a term which contributes to $A_{5;1}(\qb_1,q_2,g_3,g_4,\gamma_5)$, but
vanishes in $A_{5;1}(\qb_1,q_2,g_3,g_4,g_5)$,
\begin{equation}
\delta A_{5;1}(\qb_1,q_2,g_3,g_4,\gamma_5) =
    {1\over N_c}\left(\Tr\left[Q_{f}\right] -
       Q_qn_f\right)\sum_{\sigma\in {\rm COP}\{g_5\}\{\qb_1q_2g_3g_4\}}
       A_{5}^{[{1\over2}]}(\sigma).
\end{equation}
This term vanishes in pure QCD because all quarks carry the same color
charge.  Because they do not carry the same electric charge, the term
reappears when a gluon is changed into a photon.

In the results below, we drop all reference to any hypothetical scalar
particles.  They can be added back in by noting that scalar terms
enter in exactly the same way as the fermion terms do.  Thus, to
reinsert the scalars, simply duplicate the $A^{[{1\over2}]}$ terms
and replace $n_f\rightarrow n_s$, $\Tr\left[Q_{f}\right]
\rightarrow\Tr\left[Q_{s}\right]$ and
$A^{[{1\over2}]}\rightarrow A^{[0]}$.

The full one-loop $\overline{q}qgg\gamma$ amplitude is
\begin{eqnarray}
\label{eq:qqGGg}
{\cal A}^\oneloop(\qb_1,q_2;g_3,g_4;
   \gamma_5) &=& \sqrt{2}eg^4\left\{
            \sum_{\sigma\in Z_2}N_c
            \left(T^{\sigma_3}T^{\sigma_4}\right)^{\ib_1}_{i_2}
            A_{5;1}(\qb_1,q_2;g_{\sigma_3},g_{\sigma_4};
            \gamma_5)\right.\nonumber \\
            && \phantom{\sqrt{2}eg}+\sum_{\sigma\in Z_2}
            \Tr\left(T^{\sigma_3}\right)\left(T^{\sigma_4}
            \right)^{\ib_1}_{i_2}A_{5;2}
            (\qb_1,q_2;g_{\sigma_3};g_{\sigma_4};\gamma_5)\nonumber\\
            && \left.\phantom{\sqrt{2}eg}
            +\Tr\left(T^{3}T^{4}\right)\delta^{\ib_1}_{i_2}
            A_{5;3}(\qb_1,q_2;g_3,g_4;\gamma_5)
            \right\},
\end{eqnarray}
where the partial amplitudes are given by
\begin{eqnarray}
\label{eq:AqqGGgFULL}
A_{5;1}(\qb_1,q_2;g_3,g_4;\gamma_5) &=&
       -Q_q\left\{\vphantom{\left[{1\over N_c^2}\right]}
   A_{5}^{L}(\qb_1,g_5,q_2,g_3,g_4)+{n_f\over N_c}
   A_{5}^{[{1\over2}]}(\qb_1,g_5,q_2,g_3,g_4)\right.\nonumber\\
  &&+{1\over N_c^2}
   \left[A_5^{R}(\qb_1,q_2,g_3,g_4,g_5)
   +A_5^{R}(\qb_1,q_2,g_3,g_5,g_4)\right.\nonumber\\
  &&\left.\left.\qquad\vphantom{{n_f\over N_c}A_{5}^{[{1\over2}]}}
   +A_5^{R}(\qb_1,q_2,g_5,g_3,g_4)\right]
   \vphantom{\left[{1\over N_c^2}\right]}\right\}\nonumber\\
  &&+{\Tr\left[Q_{f}\right]\over N_c}
  \left\{A_{5}^{[{1\over2}]}(\qb_1,q_2,g_3,g_4,g_5)
       + A_{5}^{[{1\over2}]}(\qb_1,q_2,g_3,g_5,g_4)\right.\nonumber\\
  &&\phantom{{\Tr\left[Q_{f}\right]\over N_c}}
     +\left.A_{5}^{[{1\over2}]}(\qb_1,q_2,g_5,g_3,g_4)
     + A_{5}^{[{1\over2}]}(\qb_1,g_5,q_2,g_3,g_4)\right\}\nonumber\\
\\
A_{5;2}(\qb_1,q_2;g_3;g_4;\gamma_5) &=&
  Q_q\left\{\sum_{\sigma\in Z_2}A_5^{L}(\qb_1,g_5,q_2,
       g_{\sigma_3},g_{\sigma_4}) - \sum_{\sigma\in Z_2}
       A_5^{R}(\qb_1,g_4,q_2,g_{\sigma_3},g_{\sigma_5})\right\}\nonumber\\
\\
A_{5;3}(\qb_1,q_2;g_3,g_4;\gamma_5) &=&
  -Q_q\sum_{\sigma\in Z_2} \left\{
      A_5^{L}(\qb_1,g_5,q_2,g_{\sigma_3},g_{\sigma_4})
   +
    A^{R}_5(\qb_1,g_5,q_2,g_{\sigma_3},g_{\sigma_4}) \right\} \nonumber\\
  &&+ \sum_{\sigma\in S_3}\left\{
    \vphantom{{\Tr\left[Q_{f}\right]\over N_c}}
    Q_qA_5^{R}(\qb_1,g_{\sigma_3},q_2,g_{\sigma_4},g_{\sigma_5})
  + Q_qA_5^{R}(\qb_1,q_2,g_{\sigma_3},g_{\sigma_4},g_{\sigma_5})\right.
    \nonumber\\
  &&\left. \qquad\qquad\qquad
  - {\Tr\left[Q_{f}\right]\over N_c}
      A_5^{[{1\over2}]}(\qb_1,q_2,g_{\sigma_3},g_{\sigma_4},g_{\sigma_5})
  \right\}.\nonumber\\
\end{eqnarray}
The one-loop correction to $\sigma_{\overline{q}qgg\gamma}$ is
\begin{eqnarray}
\label{eq:qqGGgloop}
\sigma^\oneloop_{\overline{q}qgg\gamma} &=&
   4e^2g^6(N_c^2-1)\sum_{\rm helicities}{\rm Re}\sum_{\sigma\in Z_2}
  {A_5^{\tree}}^{*}(\qb_1,q_2;g_{\sigma_3},g_{\sigma_4};\gamma_5)
  \times\\
  &&\hspace*{-1cm}\left\{(N_c^2-1)A_{5;1}(\qb_1,q_2;g_{\sigma_3},
  g_{\sigma_4};\gamma_5)-A_{5;1}(\qb_1,q_2;g_{\sigma_4},g_{\sigma_3};
  \gamma_5) + A_{5;3}(\qb_1,q_2;g_3,g_4;\gamma_5)\right\}.\nonumber
\end{eqnarray}

\subsection{One-loop corrections to ${\qb}qg\gamma\gamma$
amplitudes}
This process has also been discussed by Signer~\cite{AS}, but in much
less detail than the ${\qb}qgg\gamma$ process.  By breaking the
permutation sum up into primitive components we are able to obtain far
more manageable (though still large) expressions.  The color decomposition
for this process is
\begin{eqnarray}
\label{eq:qqGgg}
{\cal A}^\oneloop(\qb_1,q_2;g_3;\gamma_4,\gamma_5)
    &=& {2}e^2g^3\left\{N_c\left(T^{3}\right)^{\ib_1}_{i_2}
                A_{5;1}(\qb_1,q_2;g_3;\gamma_4,\gamma_5)\right.\nonumber\\
    && \phantom{2e^2g^3}
            + \left. \Tr\left(T^{3}\right)\delta^{\ib_1}_{i_2}
            A_{5;2}(\qb_1,q_2;g_3;\gamma_4,\gamma_5) \right\} \, .
\end{eqnarray}
The primitive decompositions of the partial amplitudes are
\begin{eqnarray}
\label{eq:AVfqqGggFulla}
  A_{5;1}(\qb_1,q_2;g_3;\gamma_4,\gamma_5) &=&
  -Q_q^2\sum_{\sigma\in Z_2}
    A_{5}^{R}(\qb_1,g_3,q_2,g_{\sigma_4},g_{\sigma_5})\nonumber\\
  &&\hspace*{-3cm} -\sum_{\sigma\in S_3}\left[{Q_q^2\over N_c^2}
     A_{5}^{R}(\qb_1,q_2,g_{\sigma_3},g_{\sigma_4},g_{\sigma_5})
     -{\Tr\left[Q^2_{f}\right]\over N_c}A_5^{[{1\over2}]}
      (\qb_1,q_2,g_{\sigma_3},g_{\sigma_4},g_{\sigma_5})\right]\\
  A_{5;2}(\qb_1,q_2;g_3;\gamma_4,\gamma_5) &=&
  Q_q^2\sum_{\sigma\in Z_2}
    A_{5}^{R}(\qb_1,g_3,q_2,g_{\sigma_4},g_{\sigma_5})\nonumber\\
  &&\hspace*{-3cm} +\sum_{\sigma\in S_3}\left[Q_q^2
     A_{5}^{R}(\qb_1,q_2,g_{\sigma_3},g_{\sigma_4},g_{\sigma_5})
     -{\Tr\left[Q^2_{f}\right]\over N_c}A_5^{[{1\over2}]}
      (\qb_1,q_2,g_{\sigma_3},g_{\sigma_4},g_{\sigma_5})\right]\,,
\end{eqnarray}
where $\Tr\left[Q^2_{f}\right]$ is the sum of the
squared charges of all $n_f$ quarks.  The one-loop correction
to the cross section is
\begin{equation}
\label{eq:qqGggloop}
\sigma^\oneloop_{\overline{q}qg\gamma\gamma} =
   8e^4g^4N_c(N_c^2-1)\sum_{\rm helicities}{\rm Re}\left[
  {A_5^{\tree}}^{*}(\qb_1,q_2;g_3;\gamma_4,\gamma_5)
  A_{5;1}(\qb_1,q_2;g_3;\gamma_4,\gamma_5)\right].
\end{equation}

\subsection{One-loop corrections to ${\qb}q\gamma\gamma\gamma$
amplitudes}
This process has not been discussed in detail before.  The color
structure is trivial,
\begin{equation}
\label{eq:qqggg}
  {\cal A}^\oneloop(\qb_1,q_2;\gamma_3,\gamma_4,\gamma_5) =
  {2}\sqrt{2}e^3g^2N_c\delta^{\ib_1}_{i_2} 
   A_{5;1}(\qb_1,q_2;\gamma_3,\gamma_4,\gamma_5),
\end{equation}
with
\begin{equation}
\label{eq:Aqqggg}
A_{5;1}(\qb_1,q_2;\gamma_3,\gamma_4,\gamma_5) = Q_q^3\left(1-{1\over
   N_c^2}\right)\sum_{\sigma\in S_3}A_5^{R} 
   (\qb_1,q_2,g_{\sigma_3},g_{\sigma_4},g_{\sigma_5}).
\end{equation}

The one-loop correction to the cross section is
\begin{equation}
\label{eq:qqgggloop}
\sigma^\oneloop_{\overline{q}q\gamma\gamma\gamma} =
   16e^6g^2Q_q^6N_c^2\sum_{\rm helicities}{\rm Re}\left[
  {A_5^{\tree}}^{*}(\qb_1,q_2;\gamma_3,\gamma_4,\gamma_5)
  A_{5;1}(\qb_1,q_2,g_3,g_4,g_5)\right].
\end{equation}

\subsection{One-loop corrections to $\qb q\Qb Q\gamma$ amplitudes}
\label{sec:phot4qamps}

In processes involving four quarks, one also obtains new color factors
at one loop.  However, for the one case of interest in this paper, the
four-quark one-gluon amplitudes, the only new color structure involves
the trace of the lone gluon generator and therefore vanishes. The
authors of ref.~\cite{KST} do not provide the partial amplitude that
would have accompanied it.  However, Signer~\cite{AS} has given
the one-loop four-quark one-photon amplitudes explicitly.

The color decomposition of the four-quark, one-photon amplitudes at
one-loop is the same as at tree level, eq.~(\ref{FourQuarkGluonPhot}),
\begin{eqnarray}
\lefteqn{ {\cal A}_5^\oneloop(\qb_1,q_2,\overline{Q}_3,Q_4,\gamma_5) }
\label{FourQuarkPhotonLoop}\\
&& = \sqrt{2} e g^4  \Biggl[ \delta_{i_4}^{\ib_1}\ \delta_{i_2}^{\ib_3}\
    A_5^\oneloop(\qb_1,q_2;\overline{Q}_3,Q_4;\gamma_5) -
  {1\over N_c}\, \delta_{i_2}^{\ib_1}\ \delta_{i_4}^{\ib_3}\
    B_5^\oneloop(\qb_1,q_2;\overline{Q}_3,Q_4;\gamma_5) \Biggr]\, ,\nonumber
\end{eqnarray}
with
\begin{eqnarray}
A_5^\oneloop (\qb_1,q_2;\overline{Q}_3,Q_4;\gamma_5) &=& Q_q 
u_2^{(1)}(\qb_1,q_2;\overline{Q}_3,Q_4;\gamma_5) +
Q_Q d_2^{(1)}(\qb_1,q_2;\overline{Q}_3,Q_4;\gamma_5) \nonumber\\
B_5^\oneloop(\qb_1,q_2;\overline{Q}_3,Q_4;\gamma_5) &=& Q_q 
u_1^{(1)}(\qb_1,q_2;\overline{Q}_3,Q_4;\gamma_5) +
Q_Q d_1^{(1)}(\qb_1,q_2;\overline{Q}_3,Q_4;\gamma_5)\, .\nonumber\\
\label{abfunctions}
\end{eqnarray}
The functions $u^{(1)}_i$
\begin{eqnarray}
\lefteqn{  u_1^{(1)}(\qb_1,q_2;\overline{Q}_3,Q_4;\gamma_5) } 
\nonumber\\
&=&
ic_\Gamma\, N_c\, \left[ u_1^l(\qb_1,q_2;\overline{Q}_3,Q_4;\gamma_5)
+ {1\over N_c^2}\, u_1^s(\qb_1,q_2;\overline{Q}_3,Q_4;\gamma_5)
- {n_{\! f}\over N_c} u^{n_{\! f}}(\qb_1,q_2;\overline{Q}_3,Q_4;\gamma_5) 
\right] \nonumber\\
\label{ufunctionsa}\\
\lefteqn{ u_2^{(1)}(\qb_1,q_2;\overline{Q}_3,Q_4;\gamma_5) }\nonumber\\ 
&=&
ic_\Gamma\, N_c\, \left[ u_2^l(\qb_1,q_2;\overline{Q}_3,Q_4;\gamma_5)
+ {1\over N_c^2}\, u_2^s(\qb_1,q_2;\overline{Q}_3,Q_4;\gamma_5)
+ {n_{\! f}\over N_c} u^{n_{\! f}}(\qb_1,q_2;\overline{Q}_3,Q_4;\gamma_5) 
\right]\, \nonumber\\
\label{ufunctionsb}
\end{eqnarray}
with functions $d^{(1)}_i$ in \eqn{abfunctions} decomposed in the same
way\footnote{Note that eq.~(19) in ref.~\cite{AS} (the equivalent of our
eq.~(\ref{ufunctionsa})) is missing a factor of $N_c$~\cite{ASa}.}.
Our \eqn{FourQuarkPhotonLoop} differs from its in equivalent in
ref.~\cite{AS} because we adopt the convention of
refs.~\cite{BDK,BDKa} and define $c_\Gamma$ to be
\begin{equation}
c_{\Gamma} = {1\over 
(4\pi)^{2-\epsilon}}\, {\Gamma(1+\epsilon)\,
\Gamma^2(1-\epsilon)\over \Gamma(1-2\epsilon)}\, .\label{cgam}
\end{equation}
This differs from the convention of ref.~\cite{AS} by a factor of
${1/(16\pi^2)}$.

All of the functions on the right hand side of
\Eqns{ufunctionsa}{ufunctionsb} and the 
corresponding functions $d^{(1)}_i$ are given explicitly in ref.~\cite{AS}.
The expressions given there, however, are written in terms of symbols
which are only defined in ref.~\cite{KST}.  In general, these symbols
represent recurring combinations of logarithms and dilogarithms of the
kinematic invariants.  It turns out that these symbols are in a
one-to-one correspondence to a different set of symbols used by Bern,
Dixon and Kosower in refs.~\cite{BDK,BDKa}.  To facilitate the
understanding of these expressions, we present a conversion table in
Appendix~\ref{app:table}.

In evaluating the tree-level amplitude one finds that
\begin{equation}
B_5^{\tree}(\qb_1,q_2;\overline{Q}_3,Q_4;\gamma_5) = 
A_5^{\tree}(\qb_1,q_2;\overline{Q}_3,Q_4;\gamma_5).
\end{equation}
This degeneracy causes the
$B_5^\oneloop(\qb_1,q_2;\overline{Q}_3,Q_4;\gamma_5)$ term to drop out
of the interference term, with the result that the one-loop correction
to the cross section is 
\begin{equation}
\sigma^\oneloop_{{\qb}q{\Qb}Q\gamma}=
  4e^2g^6(N_c^2-1)\sum_{\rm helicities}{\rm Re}\left[
     {A_5^{\tree}}^{*}(\qb_1,q_2;\overline{Q}_3,Q_4;\gamma_5)
      A_5^\oneloop(\qb_1,q_2;\overline{Q}_3,Q_4;\gamma_5)\right].
\end{equation}
Thus, $B_5^\oneloop(\qb_1,q_2;\overline{Q}_3,Q_4;\gamma_5)$ need not
be computed for NLO calculations, but would be needed for inclusion in
a NNLO calculation.

\section{Conclusions}

In this paper, we have computed compact expressions for the 
six-parton tree-level amplitudes involving up to three photons.
In addition, we have presented detailed expansions for the corresponding 
five-parton one-loop amplitudes, in terms of known results.
These are all of the amplitudes needed to compute at NLO two-parton to
three-parton scattering where up to three partons are photons,
and two-parton to two-parton inclusive scattering where three partons 
are photons.
They also contribute at NNLO to two-parton to two-parton scattering where
one or two partons are photons.

\section*{Acknowledgments}

The work of W.B.K. was supported by the US Department of Energy under
grant DE-AC02-98CH10886.  W.B.K. would also like to acknowledge the
kind hospitality of the Dipartimento di Fisica Teorica at the
Universit\`a di Torino.

\appendix

\section{Two-quark, three-gluon, one-photon amplitudes}
\label{app:2q3g1gamma}

The two-quark, three-gluon, one-photon amplitudes of
\Eqns{eq:2q3g1gamma1}{eq:2q3g1gamma3}
can be computed from the two-quark, four-gluon amplitudes by summing
over all the possible positions of a gluon in the color ordering.
A more efficient procedure is to sum over the positions of a gluon
of the corresponding two-gluino, four-gluon amplitude.
The ensuing sum can then be simplified by using the dual Ward
identities~\cite{mpReview}, and can be reformulated in terms of a sum
of purely gluonic amplitudes and two-quark, four-gluon amplitudes by
using supersymmetric Ward identities~\cite{mpReview,susy}. For the
configurations of \Eqns{eq:2q3g1gamma1}{eq:2q3g1gamma3}, we obtain 
\begin{eqnarray}
\lefteqn{ A_6^\tree(\qb_5^+, q_6^-, g_1^-, g_2^-, g_3^+, \gamma_4^+) =
- \left( {\spa5.1 \over \spa6.1} + {\spa2.1 \spb4.2 \over
\spa6.1 \spb4.5} \right)\, A_6^\tree(g_1^-, g_2^-, g_3^+, g_5^+, g_4^+,
g_6^-) } \nonumber\\ &&\hspace*{3.8cm}
 -{\spa2.1 \spb4.3 \over \spa6.1 \spb4.5}\,
A_6^\tree(\qb_3^+, q_2^-, g_1^-, g_6^-, g_4^+, g_5^+)\nonumber\\
%
%
%
\lefteqn{ A_6^\tree(\qb_5^+, q_6^-, g_1^+, g_2^-, g_3^-, \gamma_4^+) =}
\nonumber\\ &&- {\spa5.2 \over \spa6.2}\, A_6^\tree(g_1^+,
g_2^-, g_3^-, g_5^+, g_4^+, g_6^-) + {\spa3.2 \over \spa6.2}\,
A_6^\tree(\qb_5^+, q_3^-, g_2^-, g_1^+, g_6^-, g_4^+) \label{2q3g1gammasub}\\
\lefteqn{ A_6^\tree(\qb_5^+, q_6^-, g_1^-, g_2^+, g_3^-, \gamma_4^+) =}
\nonumber\\ &&- {\spa5.1 \over \spa6.1}\, A_6^\tree(g_1^-,
g_2^+, g_3^-, g_5^+, g_4^+, g_6^-) + {\spa3.1 \over \spa6.1}\,
A_6^\tree(\qb_5^+, q_3^-, g_2^+, g_1^-, g_6^-, g_4^+) \,. \nonumber
%
\end{eqnarray}
All the other configurations can be obtained from the ones above by
parity inversion and/or charge conjugation.

\section{Two-quark, two-gluon, two-photon amplitudes}
\label{app:2q2g2gamma}

In converting two-quark, four-gluon amplitudes into
two-quark, two-gluon, two-photon amplitudes, we follow
the same procedure as in Appendix~\ref{app:2q3g1gamma}.
For the configurations of \tab{2q2g2gamma}, we have obtained
\begin{eqnarray}
\lefteqn{ A_6^\tree(\qb_5^+, q_6^-, g_1^-, g_2^-, \gamma_3^+,
\gamma_4^+) =}
\nonumber\\ && \sum_{\sigma\in Z_2} \left( {\spa5.1 \over
\spa6.1}\, A_6^\tree(g_1^-, g_2^-, g_5^+, g_{\sigma_3}^+,
g_{\sigma_4}^+, g_6^-) - {\spa2.1 \over \spa6.1}\, A_6^\tree(\qb_5^+,
q_2^-, g_1^-, g_6^-, g_{\sigma_4}^+, g_{\sigma_3}^+) \right)\, \nonumber\\
%
%
\lefteqn{ A_6^\tree(\qb_5^+, q_6^-, g_1^+, g_2^-, \gamma_3^-,
\gamma_4^+) =}
\nonumber\\ && {\spa5.3 \over \spa6.3}\, A_6^\tree(g_1^+, g_2^-, g_5^+, g_4^+,
g_3^-, g_6^-) + {\spa5.2 \over \spa6.2}\, A_6^\tree(g_1^+, g_2^-, g_5^+,
g_3^-, g_4^+, g_6^-) \nonumber\\ && - {\spa2.3 \over \spa6.3}\,
A_6^\tree(\qb_5^+, q_2^-, g_1^+, g_6^-, g_3^-, g_4^+) -
{\spa3.2 \over \spa6.2}\,
A_6^\tree(\qb_5^+, q_3^-, g_4^+, g_6^-, g_1^+, g_2^-) \\
\lefteqn{ A_6^\tree(\qb_5^+, q_6^-, g_1^-, g_2^+, \gamma_3^-,
\gamma_4^+) =}
\nonumber\\ && \left( {\spa5.1 \over \spa6.1} + {\spa3.1 \spb2.3 \over
\spa6.1 \spb2.5} \right)\, A_6^\tree(g_1^-, g_2^+, g_5^+, g_4^+,
g_3^-, g_6^-) + {\spa5.1 \over \spa6.1}\, A_6^\tree(g_1^-, g_2^+, g_5^+,
g_3^-, g_4^+, g_6^-) \nonumber\\ && + {\spa3.1 \spb2.4 \over \spa6.1\spb2.5}\,
A_6^\tree(\qb_4^+, q_3^-, g_6^-, g_1^-, g_2^+, g_5^+) -
{\spa3.1 \over \spa6.1}\,
A_6^\tree(\qb_5^+, q_3^-, g_4^+, g_6^-, g_1^-, g_2^+) \,,\nonumber
\end{eqnarray}
where the sum in the first equation is over permutations of 3 and
4. All the other configurations can be obtained from the ones above by
photon relabeling, parity inversion and/or charge conjugation.

\section{Translating symbols used in NLO matrix elements}
\label{app:table}
The matrix elements for the four-quark one-photon process at one-loop
are given in ref.~\cite{AS} in terms of symbols which are defined in
ref.~\cite{KST}. These symbols have a one-to-one correspondence with
symbols defined in ref.~\cite{BDK} that are used in the two-quark
three-gluon primitives.  To facilitate the reading and use of
ref.~\cite{AS}, we provide a translation table for those symbols.

\begin{table}[ht]
\addtolength{\arraycolsep}{0.1cm}
\renewcommand{\arraystretch}{1.4}
$$
\begin{array}{c|l}
\hline\hline
  \mbox{Signer (ref.~\cite{AS})} 
&\qquad \mbox{Bern,{\it et. al.\/} (ref.~\cite{BDK}) } \\
\hline
 \displaystyle \L0{^{i j}_{k l}} 
&\displaystyle\qquad \ln \left(\frac{-s_{ij}}{-s_{kl}}\right) \\[4pt]
 \displaystyle \L1{^{i j}_{k l}} 
&\displaystyle\qquad \frac{1}{s_{kl}} \mbox{L}_0\left(\frac{-s_{ij}}{-s_{kl}}\right)\\[4pt]
 \displaystyle \L2{^{i j}_{k l}} 
&\displaystyle\qquad \frac{1}{s_{kl}^2} \mbox{L}_1\left(\frac{-s_{ij}}{-s_{kl}}\right)\\[4pt]
 \displaystyle \L3{^{i j}_{k l}} 
&\displaystyle\qquad \frac{1}{s_{kl}^3} \mbox{L}_2\left(\frac{-s_{ij}}{-s_{kl}}\right)\\[4pt]
 \displaystyle {\cal F}(i,j,k) 
&\displaystyle\qquad \left.\mbox{Ls}_{-1}\left(\frac{-s_{ij}}{-s_{mn}},
 \               \frac{-s_{jk}}{-s_{mn}}\right)\right|_{
                  \{m,n\}=\{1,2,3,4,5\}/\{i,j,k\}} \\[4pt]
 \displaystyle {\cal P}_{ij} 
&\displaystyle\qquad \left(\frac{\mu^2}{-s_{ij}}\right)^\epsilon\\[4pt]
 \displaystyle \langle ijkl \rangle   
&\displaystyle\qquad \frac{\spa{i}.k \spa{j}.l}{\spa{i}.l \spa{j}.k}\\[7pt]
\hline\hline
\end{array}
$$
\caption{Conversion table relating the undefined symbols in
ref.~\cite{AS} to their equivalents defined in ref.~\cite{BDK}}
\label{tab:AS-BDK}
\end{table}

\vfil\eject

\end{document}